\documentclass[
    ,final            
  ]
  {aipproc}

\layoutstyle{8x11single}

\usepackage{amsmath,amsfonts,amssymb,dsfont,mathrsfs,pgfplots,tikz,graphicx}

\begin{document}

\title{White dwarfs constraints on dark sector models with light particles}


\author{Lorenzo Ubaldi}{
  address={Physikalisches Institut der Universit\"{a}t Bonn, Nussallee 12, D-53115 Bonn, Germany}
}

\begin{abstract}
The white dwarf luminosity function is well understood in terms of standard model physics and leaves little room for exotic cooling mechanisms related to the possible existence of new weakly interacting light particles. This puts significant constraints on the parameter space of models that contain a massive dark photon and light dark sector particles.     
\end{abstract}

\maketitle


\section{Introduction}

A white dwarf (WD) is a simple astronomical object. It represents the final stage of evolution of a star whose initial mass was up to $\sim 8 \ M_\odot$, with $M_\odot$ the solar mass. WDs are not burning nuclear fuel any longer, they are simply cooling down. Their internal temperature is of the order of keV. The WD core is composed of ions of carbon and oxygen, and of highly degenerate electrons. It is the degeneracy pressure of the electrons that supports the star and prevents it from collapsing under its self gravitational pull. WDs can be classified as DA and non-DA. The residual star's atmosphere is hydrogen rich for the first category, helium rich for the second. Observations indicate that 85\% of WDs are DA~\cite{Althaus:2010pi}.

\begin{figure}[!t]
\centering
\includegraphics[width=0.7\textwidth]{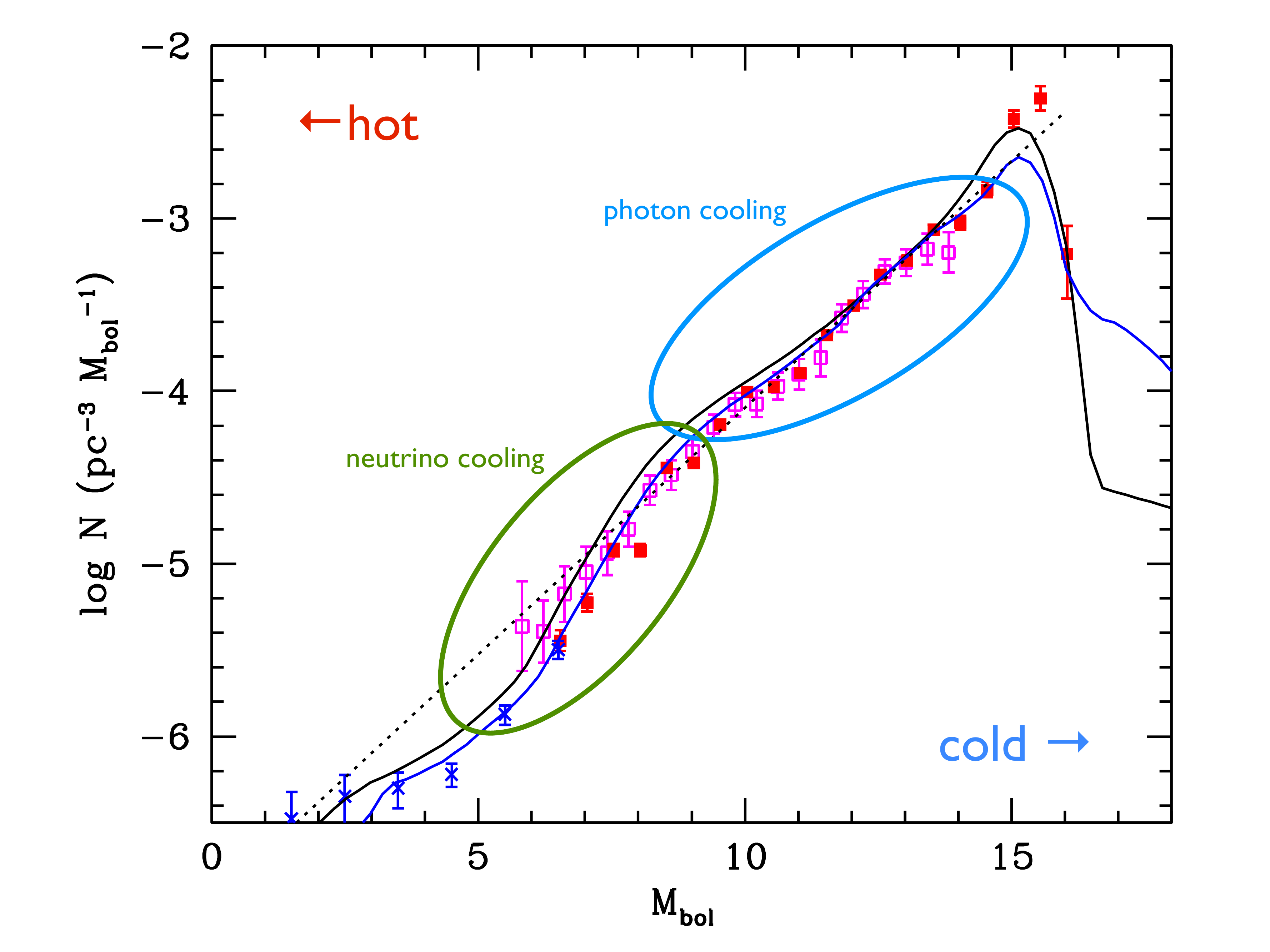}
\caption{\textit{Luminosity function of white dwarfs.}  Red (Harris \textit{et al.}~\cite{Harris:2005gd}) and blue (Krzesinski \textit{et al.}~\cite{Krzesinski:2009}) points represent the luminosity 
function of all white dwarfs (DA and non-DA families).  Magenta points~\cite{DeGennaro:2007yw} represent the luminosity function of the DA 
white dwarfs alone. The dotted line represents the luminosity 
function obtained assuming Mestel's approximation.  The continuous lines correspond to full simulations assuming a constant star formation rate 
and an age of the Galaxy of 13 Gyr for the DA family (black line) and all, DA and non-DA, white dwarfs (blue line).}
\label{FigLF}
\end{figure}

The quantity of interest for understanding the cooling of WDs is the luminosity function (LF), defined as the number of WDs of a given luminosity, $L$, or bolometric magnitude\footnote{$L_\odot$ is the solar luminosity.}, $M_{\rm bol} = -2.5 \log_{10}(L/L_\odot ) +4.74$, per unit of magnitude interval and unit volume. A simple way of thinking about the LF is given by the following equation~\cite{Raffelt:1996wa}
\begin{equation}
\frac{dN}{dM_{\rm bol}}=-B \frac{dU}{dM_{\rm bol}}\frac{1}{L_\gamma+L_\nu + L_x},
\end{equation}
where $N$ is the number of WDs per unit volume, $B$ the WD birthrate per unit volume, $U$ the internal energy of the star, $L_\gamma, L_\nu , L_x$ the energy loss rates\footnote{The energy loss rates $L$ have units of energy per time. In the next section we use $\epsilon$ with units of energy per time per unit mass. The relation between the two is given by $L = \epsilon M_{\rm WD}$, with $M_{\rm WD}$ the WD mass.} due to the emission of photons, neutrinos and possible exotic new particles, respectively. As I explain in more detail in the next section, photon cooling affects mostly the colder WDs, and neutrino cooling is dominant in the hotter WDs. The LF obtained from observational data is shown in Fig.~\ref{FigLF}, along with curves that result from numerical simulations of the stellar evolution. The simulations only include standard model physics ($L_x = 0$) and we clearly see that they fit the data very well. This means that WDs are very well understood in terms of known microscopic physics and at the same time they provide an excellent laboratory to constrain new models of particle physics that would predict the presence of exotic cooling. 

In what follows I first describe the standard cooling mechanisms in WDs, then explore the constraints on models in which new light particles could be pair produced in the star interior. This manuscript is based on work I have done with my collaborators Dreiner, Fortin and Isern~\cite{Dreiner:2013tja}.

\section{Cooling mechanisms}

\subsection{Standard}
The carbon and oxygen ions in the WD interior form to a good approximation a classical Boltzmann gas, which stores most of the thermal energy. Through the surface layers the heat is transferred  to the exterior. Using Mestel's approximation~\cite{1952MNRAS.112..583M} one can relate the rate of energy loss at the surface to the interior temperature:
\begin{equation} \label{epsgamma}
\epsilon_\gamma=3.29\times10^{-3}\ T_7^{7/2}\ {\rm erg}\ {\rm g}^{-1}\ {\rm s}^{-1},
\end{equation}
where $T_7\equiv\frac{T}{10^7 \ {\rm K}}$. We refer to this mechanism as photon cooling.

\begin{figure}[!t]
\centering
\includegraphics[width=0.9\textwidth]{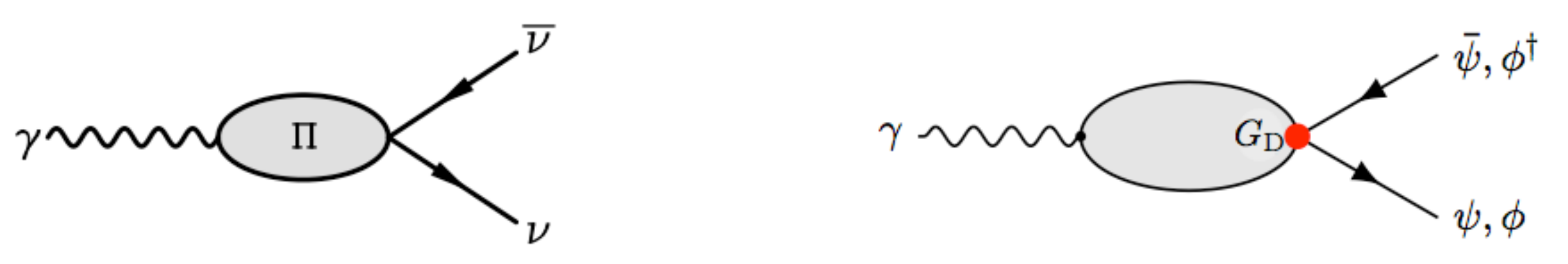}
\caption{\textit{Plasmon decays.}  The diagram on the left shows the plasmon decay into a pair of neutrinos, which results in the main standard cooling mechanism for hot white dwarfs ($M_{\rm bol} \lesssim 7$). The diagram on the right shows the analogous decay into dark sector particles, with $G_{\rm D}$ the four-fermion coupling of Fig.~\ref{Figepsi}. }
\label{FigDark}
\end{figure}

For WDs hotter than $4-5 \times 10^7$ K, that corresponds to $M_{\rm bol} \sim 6-7$, the heat is more efficiently lost via the emission of neutrinos rather than photons. The main process responsible for producing neutrinos inside WDs is the plasmon decay, depicted on the left in Fig.~\ref{FigDark}. A plasmon is a photon that, propagating in a medium, acquires also a longitudinal polarization. One could then loosely speak of a massive photon that decays into a pair of neutrinos. A more proper description of the phenomenon is the following. The electromagnetic wave propagating in the interior of the star stimulates an organized oscillation of the electrons, which in turn emit neutrino pairs via the weak interactions. The calculation of the energy loss related to the plasmon decay is involved and cannot be done analytically. One can use the semi-analytic result derived by Haft, Raffelt and Weiss~\cite{Haft:1993jt}
\begin{equation}\label{epsnu}
\epsilon_\nu=1.40\times10^{15} \ \lambda^9\gamma^6e^{-\gamma}(f_T+f_L)\ 
{\rm erg}\ {\rm g}^{-1}\ {\rm s}^{-1},
\end{equation}
where
\begin{align}
\lambda &=1.69\times10^{-3}\ T_7,\quad\quad\gamma=\frac{28}{T_7},\\
f_T &=2.4+0.6\gamma^{1/2}+0.51\gamma+1.25\gamma^{3/2},\quad\quad f_L=\frac{8.6\gamma^2+1.35\gamma^{7/2}}{225-17\gamma+\gamma^2}.
\end{align}

\subsection{Exotic}
On top of the standard mechanisms described above, particle physics beyond the standard model may imply the existence of new weakly interacting particles that could be produced in WDs and escape, contributing to their cooling. A well studied example is provided by the DFSZ axion~\cite{Dine:1981rt, Zhitnitsky:1980tq}. The main process responsible for its production would be the axion bremsstrahlung from an electron Coulomb-scattering off of a nucleus. I will not discuss this further here, see Refs.~\cite{Isern:2008fs, Isern:2012ef} for recent work on the subject.

In this work I am interested in light particles that couple to electrons but can only be pair produced, as opposed to the axion that can be produced singly. In the case of pair production the process relevant to WD cooling is the plasmon decay, as depicted on the right of Fig.~\ref{FigDark}. The energy loss rate associated with this process can easily be related to the one of eq.~\eqref{epsnu}:
\begin{equation} \label{epsx}
\epsilon_x=1.40\times10^{15}\left(\frac{C_{\rm D} G_{\rm D}}{C_VG_\text{F}}\right)^2\ \lambda^9\gamma^6e^{-\gamma}(f_T+f_L)\ 
\text{erg}\ \text{g}^{-1}\ \text{s}^{-1}.
\end{equation}
Here, $G_F = 1.166 \times 10^{-5} \ {\rm GeV}^{-2}$ is the Fermi constant, $C_V = 0.964$ is the effective neutral-current vector coupling constant, $C_{\rm D}$ is a constant of order 1 analogous to $C_V$, and $G_{\rm D}$ is the four-fermion interaction between electrons and new light states, analogous to $G_F$.

\begin{figure}[!t]
\centering
\includegraphics[width=0.9\textwidth]{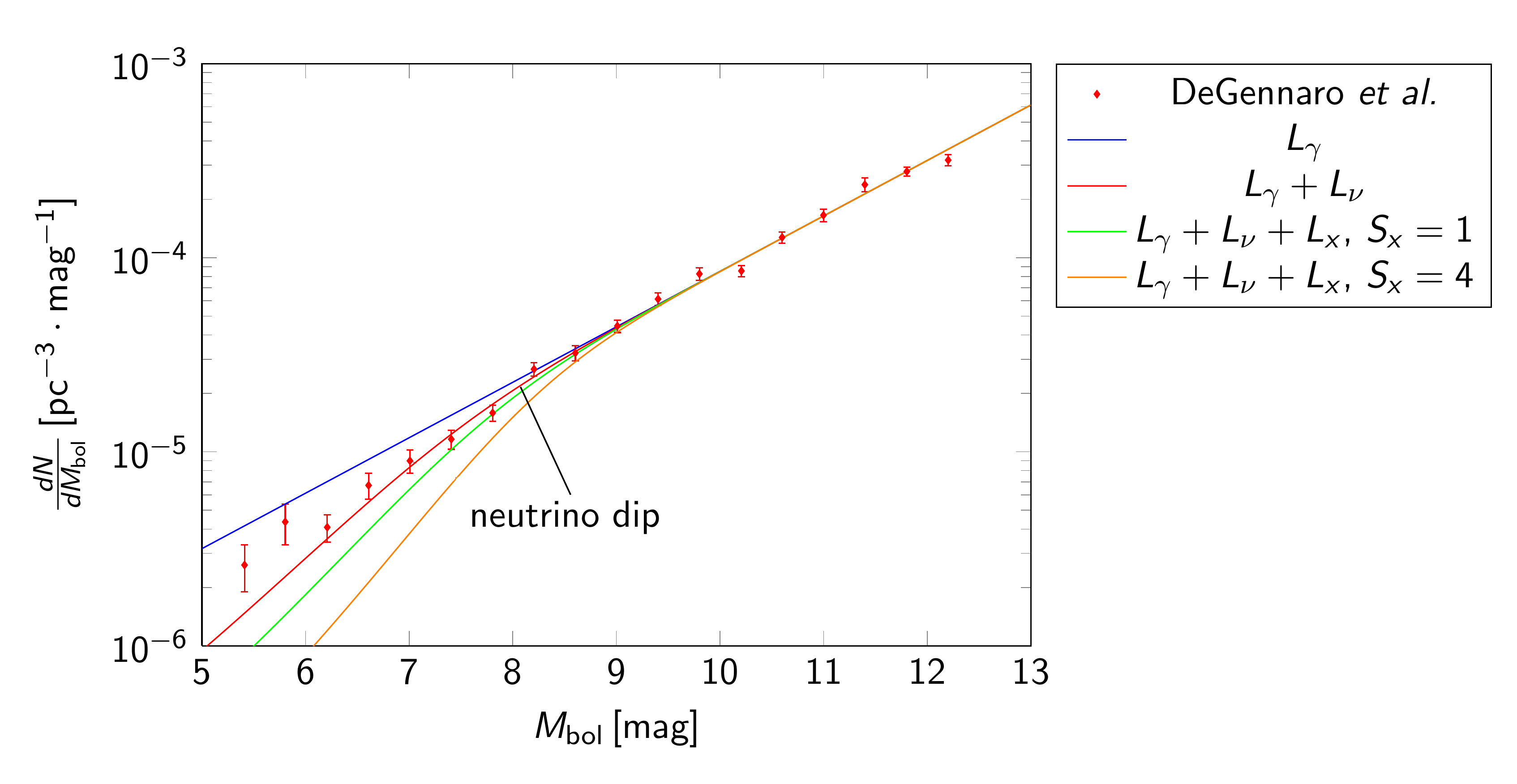}
\caption{\textit{Luminosity function with extra cooling from additional plasmon decays.} The data points shown in red are from Ref.~\cite{DeGennaro:2007yw}. The energy loss rates $L_\gamma, L_\nu, L_X$ are computed from eqs.~\eqref{epsgamma}, \eqref{epsnu}, \eqref{epsx} respectively. }
\label{Sx}
\end{figure}

It is useful to define
\begin{equation}
S_x \equiv \frac{L_x}{L_\nu} = \frac{\epsilon_x}{\epsilon_\nu} = \left(\frac{C_{\rm D} G_{\rm D}}{C_VG_\text{F}}\right)^2 \ .
\end{equation}
From Fig.~\ref{Sx} one sees that when $S_x > 1$ the extra cooling would steepen the LF too much and be in disagreement with the data. In Ref.~\cite{Dreiner:2013tja} we made this argument more precise using a $\chi^2$ fit and obtained the same constraint. Thus, models with $C_{\rm D} G_{\rm D} > C_VG_\text{F}$ are excluded. In the next section I introduce a class of models for which this constraint can be quite severe.

\section{A dark sector model}
\begin{figure}[!t]
\centering
\includegraphics[width=0.9\textwidth]{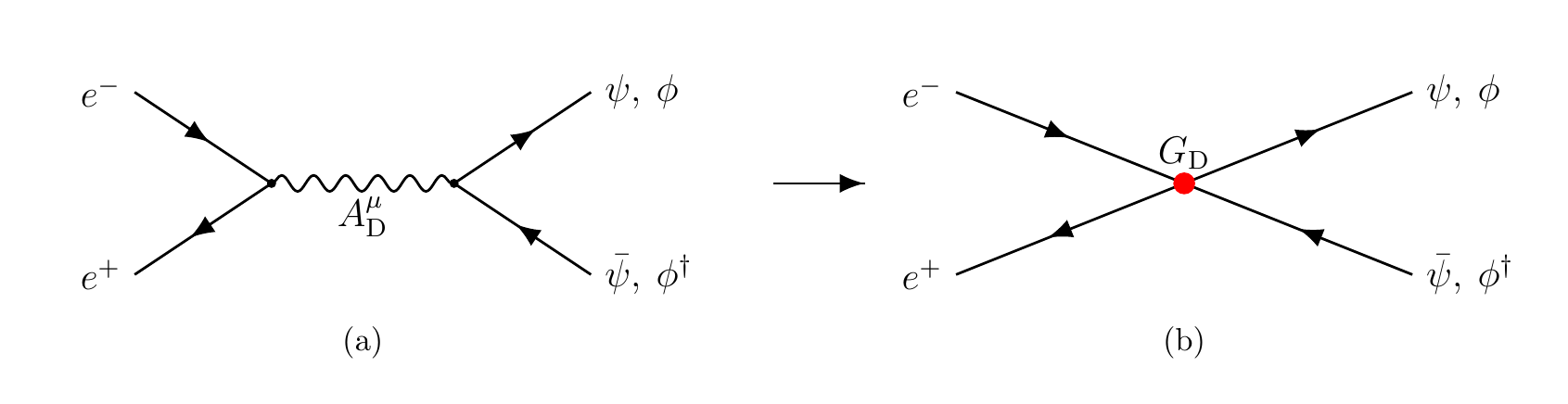}
\caption{\textit{Integrating out the dark photon.}  The red dot in the diagram on the right denotes a  dimension 6 operator.}
\label{Figepsi}
\end{figure}

Consider adding to the Standard Model (SM) an extra gauge group $U(1)_{\rm D}$ which is spontaneously broken so that the corresponding gauge boson $A^\mu_{\rm D}$ is massive~\cite{Holdom:1985ag, Fayet:1990wx}. Generically $A^\mu_{\rm D}$ mixes with the hypercharge boson $B^\mu$, so the lagrangian reads
\begin{equation}
\mathscr{L}=\mathscr{L}_\text{SM}+\mathscr{L}_\text{D}+\mathscr{L}_{\text{SM}\otimes\text{D}},\hspace{1cm}\text{where}\hspace{1cm}\mathscr{L}
_{\text{SM}\otimes\text{D}}=\frac{\varepsilon_Y}{2}B_{\mu\nu}F_\text{D}^{\mu\nu},
\end{equation}
with $B_{\mu\nu} \equiv \partial_\mu B_\nu - \partial_\nu B_\mu$,  $F_\text{D}^{\mu \nu} \equiv \partial^\mu A_\text{D}^\nu - \partial^\nu A_\text{D}^\mu$, and $\varepsilon_Y$ is a small mixing parameter. Below the electroweak scale the kinetic mixing is between the dark photon and the SM photon, $A^\mu$, with mixing parameter $\varepsilon = \varepsilon_Y \cos \theta_W$, where $\theta_W$ is the weak mixing angle.
 Let's assume that the dark sector contains fermions charged under $U(1)_{\rm D}$,
\begin{equation}
\mathscr{L}_\text{D} \supset \sqrt{4\pi \alpha_{\rm D}} \  C_{\rm D} \ A^\mu_{\rm D} \ \bar \psi \gamma_\mu \psi \ ,
\end{equation}
where $\alpha_{\rm D}$ is the dark fine structure constant. 
Rotating the gauge fields to a basis where their kinetic terms are diagonal one finds that the dark photon couples to the electromagnetic current
\begin{equation}
\mathscr{L} \supset -\varepsilon \ e \ J_\mu^\text{SM}A_\text{D}^\mu \ .
\end{equation}
This coupling is small because suppressed by $\varepsilon$. Now we see that the diagram shown on the left in Fig.~\ref{Figepsi} is possible. We consider a dark photon with a mass heavier than 1 MeV, which is high compared to the WD temperature (order keV). Thus we can integrate it out, as shown on the right in Fig.~\ref{Figepsi}, and obtain the four-fermion interaction
\begin{equation}
-G_\text{D} C_{\rm D}(\bar{\psi}\gamma^\mu\psi) (\bar{e}\gamma_\mu e),
\end{equation}
with $G_{\rm D} = \frac{4\pi\varepsilon\sqrt{\alpha\alpha_\text{D}}}{m_{A_\text{D}}^2}$ and $\alpha = \frac{e^2}{4\pi}$ the fine structure constant. 

\subsection{Constraints}
If the dark fermion $\psi$ is lighter than a few keV, then the plasmon decay depicted on the right in Fig.~\ref{FigDark} is kinematically accessible in the WD interior. In this case the constraint discussed above applies and we have that $C_{\rm D} G_{\rm D} > C_V G_F$ is excluded. This is not the end of the story. If $C_{\rm D} G_{\rm D}$ is too big, the dark fermions interact too strongly with the electrons in the WD and they get trapped within the star. When that happens we cannot apply the simple energy loss argument. We estimated in Ref.~\cite{Dreiner:2013tja} that one must satisfy the condition $C_{\rm D} G_{\rm D} < 400 \ C_V G_F$ in order to avoid trapping. 

\begin{figure}[!t]
\centering
\includegraphics[width=0.8\textwidth]{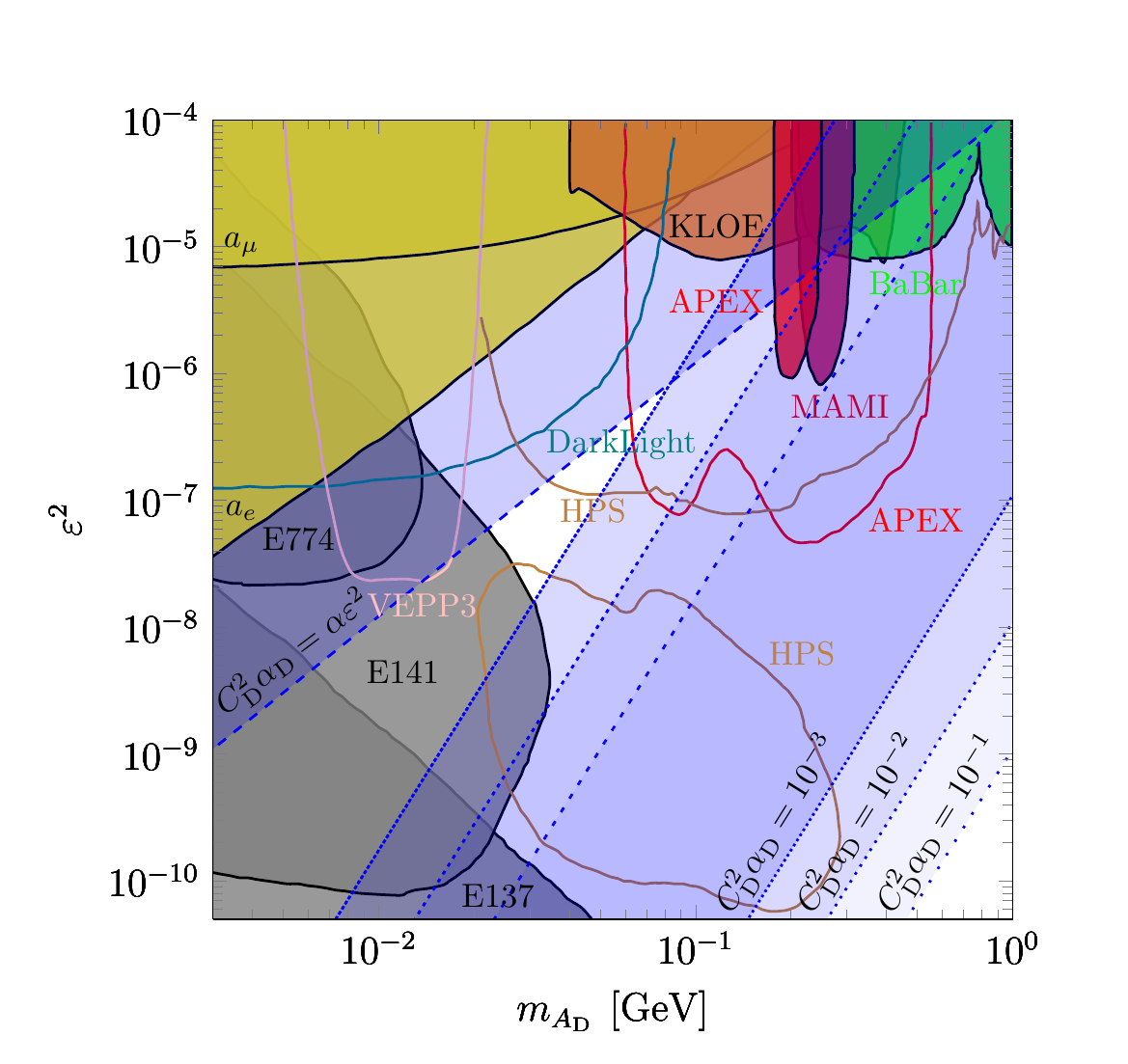}
\caption{\textit{Parameter space exclusion of dark forces with light ($\lesssim$ few tens of keV) hidden sector particles from energy losses in WDs.}  $m_{A_{\rm D}}$ is the dark photon mass, $\varepsilon$ the mixing parameter. The blue shaded regions are excluded, under the assumption that the dark fermion $\psi$ is lighter than a few keV, by WD cooling for $C_{\rm D}^2\alpha_{\rm D}=10^{-1}$ (loosely dotted lines), $C_{\rm D}^2\alpha_{\rm D}=10^{-2}$ (dotted lines) and $C_{\rm D}^2\alpha_{\rm D}=10^{-3}$ (densely dotted lines). On the left of these blue bands the hidden sector particles would be trapped inside the WD, which is why we cannot exclude that region with the simple cooling argument.  For experiments, which usually assume the dark photon decay is predominantly into the SM, shaded regions correspond to completed direct searches while curves show future reach.  For the electron and muon anomalous magnetic moments, shaded regions are excluded by measurements~\cite{Pospelov:2008zw,Davoudiasl:2012ig,Endo:2012hp}.  The bounds shown are from beam dump experiments at SLAC: E137, 
E141 
and E774~\cite{Riordan:1987aw,Bross:1989mp,Andreas:2012mt}; $e^+e^-$ colliding experiments: BaBar~\cite{Aubert:2009au,Bjorken:2009mm} 
and KLOE~\cite{Archilli:2011zc}; and fixed-target experiments: APEX~\cite{Abrahamyan:2011gv}, DarkLight~\cite{Freytsis:2009bh}, HPS~
\cite{Boyce:2012ym}, MAMI~\cite{Merkel:2011ze} and VEPP-3~\cite{Wojtsekhowski:2012zq}. 
}
\label{FigExclusion}
\end{figure}

Altogether, under the assumption that $\psi$ is lighter than a few keV, we can exclude the following region of the parameter space
\begin{equation}
1.09\times10^{-10}\left(\frac{m_{A_\text{D}}}{\text{GeV}}\right)^4=\frac{C_V^2G_\text{F}^2m_{A_\text{D}}^4}{16\pi^2\alpha}\lesssim C_\text{D}^2\alpha_\text{D}\varepsilon^2\lesssim\frac{400^2C_V^2G_\text{F}^2m_{A_\text{D}}^4}{16\pi^2\alpha}=1.09\times10^{-10}\left(\frac{20m_{A_\text{D}}}{\text{GeV}}\right)^4 \ .
\end{equation}
This constraint is shown in Fig.~\ref{FigExclusion} for different values of the free parameters $C_\text{D}^2\alpha_\text{D}$. We also show limits from laboratory experiments that have been performed, and forecasted limits from planned experiments. Several of these bounds assume that there is no content in the dark sector and that the dark photon decays into $e^+ e^-$. The reader is referred to the literature quoted in the caption of the figure for such details.

\section{Conclusions}
WDs can put significant constraints on the parameter space of dark sector models. As one can see in Fig.~\ref{FigExclusion}, these bounds are competitive with, and often complementary to, those obtained from laboratory experiments. 
 

\begin{theacknowledgments}
I would like to thank my collaborators Herbi Dreiner, Jeff Fortin and Jordi Isern with whom I wrote the original paper on this subject. I want to congratulate the organizers of the 2013 summer CETUP* program for their excellent job.  
I acknowledge the DFG SFB TR 33 ``The Dark Universe'' for support throughout this work.
 \end{theacknowledgments}



\bibliographystyle{aipproc}   

\bibliography{WD}

\IfFileExists{\jobname.bbl}{}
 {\typeout{}
  \typeout{******************************************}
  \typeout{** Please run "bibtex \jobname" to optain}
  \typeout{** the bibliography and then re-run LaTeX}
  \typeout{** twice to fix the references!}
  \typeout{******************************************}
  \typeout{}
 }

\end{document}